\documentclass[aps,prb,preprint,groupedaddress,endfloats*,amsmath]{revtex4}

\usepackage[dvips]{graphicx,color}
\usepackage{txfonts}
\newcommand{\diff}{\mathrm{d}}

\makeatletter
\def\@dotsep{4.5}
\makeatother

\begin{document}



\title{Core-shell structures in single flexible-semiflexible block
copolymers: Finding the free energy minimum for the folding transition}


\author{Natsuhiko Yoshinaga}
\affiliation{Department of Physics, Graduate School of Sciences, the
University of Tokyo, Tokyo 113-0033, Japan}

\author{Kenichi Yoshikawa} 
\affiliation{Department of Physics, Graduate School of Sciences, Kyoto University, Kyoto 606-8502, Japan}


\date{\today}

\begin{abstract}
We investigate the folding transition of a single diblock copolymer
 consisting of a semiflexible and a flexible block.
We obtain a {\it Saturn-shaped} core-shell
 conformation in  the folded state, in which the flexible block
 forms a core and the semiflexible block wraps around it.
We demonstrate two distinctive features of the core-shell structures: 
(i) The kinetics of the folding transition in the copolymer are
 significantly more efficient than those of
 a semiflexible homopolymer.
(ii) The core-shell structure does not depend on the transition pathway.

\end{abstract}

\pacs{}

\maketitle


\section{Introduction}
Polymer-based nanostructures have been extensively studied due to their
importance in industrial applications, particularly in nanodevices and
nanomachines \cite{rodrigues:2005,hamley:2003,munk:1998}.
They are also of importance in living cells; DNA and protein molecules
have nano-ordered structures, which show a close relationship with their biological functions.
Macromolecules in biological systems typically undergo conformational
transitions.
In polymer physics, this has been discussed as the coil-globule
transition \cite{lifshitz:1978}, in which a flexible swollen homopolymer collapses into
a spherical globular conformation.
The globule state is liquid-like and disordered, whereas a lot of
macromolecules, particularly biomacromolecules, have ordered, folded structures.
For this reason, there has recently been much attention to semiflexible
homopolymers, which show bending rigidity along the chain, and, as a result, have a rod-like
properties, although their contour lengths are long enough to exhibit overall fluctuations.
A semiflexible homopolymer has been extensively studied as a model for a DNA
molecules \cite{Marko:1995,ubbink:1996,sakaue:2002}.
In these works, nano-ordered structures, such as toroids, cylinders, and
rackets were investigated experimentally, theoretically and computationally \cite{noguchi:1998}.
Due to these extensive studies, it was found that the formation of nano-ordered
conformations in DNA is well reproduced in semiflexible polymers with
homogeneous bending rigidity.
In living systems, however, biopolymers such as proteins are, in general, heteropolymers with complicated
sequences of amino acids.
It is often mentioned that sequences within proteins are relevant for
their conformations in the folded states.
Therefore, most of studies heretofore conducted have dealt with sequences in
heteropolymers which lead to diversity
in conformations.
Nevertheless, our understanding of role of heterogeneity in bending
rigidity is still primitive.
In this article, we propose a minimum model to extend the concept of semiflexible
polymers toward heteropolymers.
A single block copolymer is the simplest extension of
a single homopolymer in the direction of single heteropolymers such as proteins.
To this end, we concentrate on a simple model: diblock copolymers in which two
blocks possess different levels of flexibility.


Assemblies of rod-coil copolymers have been first discussed in the
context of polymer
blends \cite{hashimoto:1986,halperin:1989b}; they have recently been
investigated in the context of polymer solutions\cite{halperin:1990}, with regard to their
applications in the field of nanocapsules \cite{jenekhe:1998}.
However, less attention has been paid to the properties of single
rod-coil copolymers.
In computer simulations, several novel structures were found by
Cooke et al. for multi-block rod-coil copolymers, which they refer to as
semiflexible copolymers.
The static properties of these species were studied \cite{cooke:2004}.
In the present study, we focus instead on the kinetics of the process of
folding into ordered phases in single rod-coil
diblock copolymers.

The kinetics and pathways of conformational transitions are relevant
since it is well known that in experiments the results strongly depend on
the method of preparation.
For example, the authors demonstrated that the
final structure in the folding transition of DNA depends on whether
condensing agents or monovalent salts are added first \cite{kidoaki:1996}.
This result suggests that we must discuss not only static
stability but also kinetics pathways for the folding transition.
Along these lines, we have investigated the dependence of folded
structures of semiflexible polymers depend on kinetic pathways \cite{noguchi:1998,yoshinaga:2005}.
Once a chain falls into a metastable state, it cannot escape to the most stable state under thermal fluctuations.
For semiflexible polymers, even when the toroidal conformation is
the most favorable,
a small hairpin-like conformation in the early stage of the folding
transition results in the formation of a cylindrical structure as a
stable state.
Due to a high free-energy barrier, the cylinder does not make the
transition to a toroid over a practical time scale.
Our interest is in the kinetics of formation of specific nano-ordered structures, and
particularly on the pathways to folded states.
We demonstrate that our single flexible-semiflexible copolymer avoids
local free energy minima and forms a
core-shell structure.

\section{Simulations}   
We carried out Langevin dynamics simulations for a single rod-coil block copolymer.
A rod-coil copolymer consists of a flexible and a semiflexible block.
Although rigid rods with infinitely large bending elasticity are often used for the rod block, in order
to allow more general discussions, we used a semiflexible polymer with bending rigidity.
We adapted a bead-spring model with the following potentials:
\begin{eqnarray} 
V_{\mathrm{beads}} 
&=& 
\frac{k}{2} \sum _{i} ({| \mathbf{r} _{i+1} -  \mathbf{r} _{i}|} - a)^2,\label{V_beads}\\
V_{\mathrm{bend}} 
&=&
\sum_{\alpha} \kappa _{\alpha} \sum _{i \in \alpha}  (1 - \cos \theta _{i} ),\\
V_{\mathrm{LJ}} 
&=& 
4 \sum_{\alpha, \beta} \epsilon _{\alpha \beta} 
\sum _{i \in \alpha,j \in \beta} 
\left[ \left( \frac{a}{|\mathbf{r} _i -
 \mathbf{r} _j|} \right)^{12} - \left( \frac{a}{|\mathbf{r} _i - \mathbf{r}
 _j|} \right)^{6} \right],
\label{V_lj}
\end{eqnarray} 
where $\mathbf{r} _{i}$ is the coordinate of the $i$th monomer and $\theta
_{i}$ is the angle between adjacent bond vectors.
The subscripts $\alpha$ and $\beta$ denote ``s'' (semiflexible) or ``f'' (flexible) monomers.
The monomer size $a$ was chosen as the unit length, and
$k_{B} T$ as the unit energy. 
The excluded volume and short-ranged attractive interactions
between monomers are included with the Lennard-Jones potential
with the coupling constant $\epsilon _{\alpha \beta}$, which determines
the strength of the attractive interaction between monomers in the $\alpha$ and $\beta$ states.
We set the spring constant to be $k = 400$.
The bending elasticity for a flexible block was chosen to be $\kappa _{f} = 0$.
We note that the persistence length of a semiflexible block can
be written as $l_p T = \kappa _{s} a$.
The length of each block was chosen as $N_{s} = N_{f} = 128$.

The equation of motion is written as,
\begin{equation}
m \frac{{\mathrm{d}}^2 {\mathbf{r}} _i}{{\mathrm{d}} t^2} 
= - \gamma \frac{{\mathrm{d}} {\mathbf{r}} _i}{{\mathrm{d}} t}
- \frac{\partial V }{\partial {\mathbf{r}} _i}
+ {\boldsymbol{\xi}} _i,
\label{simulation}
\end{equation}
where $m,\gamma $ are the mass and friction constant of monomeric units, respectively.
The unit time scale is $\tau = \gamma a^2/k_{B}T$.
We set the time step as $0.01 \tau$, and use $m=1.0$ and $\gamma =1.0$.
With these parameters, the relaxation time of the momentum of a monomer is
sufficiently fast compared to the time scale of interest.
Gaussian white noise ${\boldsymbol{\xi _{i}}}$ satisfies a 
fluctuation-dissipation relation,
\begin{equation}
 <{\boldsymbol{\xi}} _i (t) \cdot {\boldsymbol{\xi}} _j (t')> = 6 \gamma
  k_{B}T \delta _{ij} \delta (t-t').
\end{equation}

\section{Core-shell structure}
\begin{figure}
\includegraphics{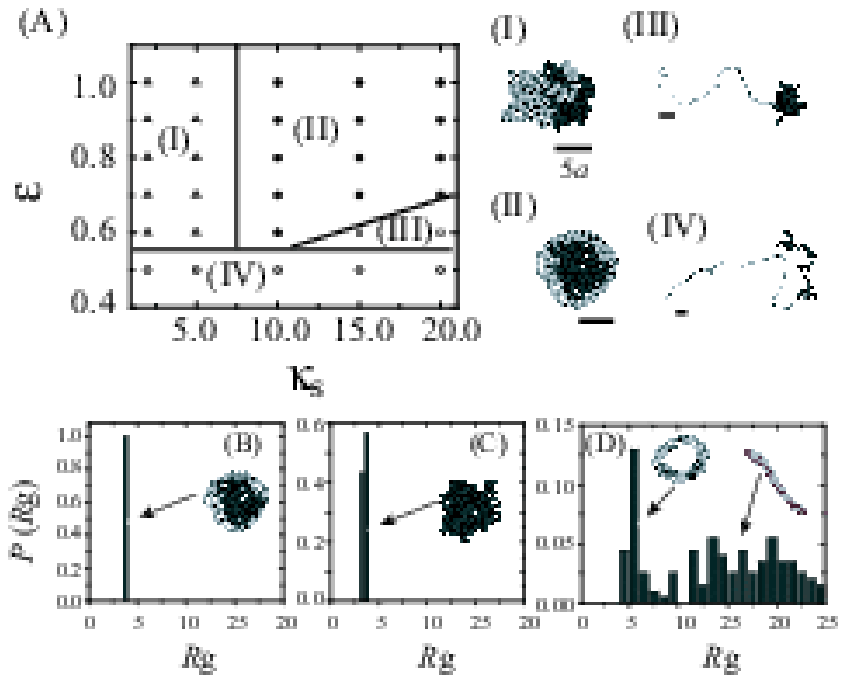}
\caption{Phase diagram of single folded rod-coil copolymers at
 various $\epsilon$ and $\kappa _{s}$ (A), and illustrations of the
 conformations the polymer in each phase: (I) disordered collapse, (II) {\it Saturn-shape} phase
 separation, (III) {\it tadpole}, and (IV) swollen. The probability
 distribution of the gyration radius of a rod-coil copolymer (B) and a
 flexible (C) and semiflexible (D) homopolymer in the collapsed state
 are also shown.
In the figures, the monomer in the flexible block is shown in light
 gray, while the monomer in the semiflexible block is shown in dark gray. 
A semiflexible homopolymer has two stable conformations, toroidal and cylindrical.
\label{fig:1}
}
\end{figure}

First, we consider the system in non-selective solvents, i.e. $\epsilon _{ss} = \epsilon _{sf} = \epsilon _{ff} = \epsilon$.
Figure 1 shows typical conformations of folded single
rod-coil block copolymers.
At sufficiently large $\epsilon$, a rod-coil copolymer in the elongated
state undergoes a transition to the folded state.
For small value of $\kappa _{s}$, the collapse is disordered, while a {\it Saturn-shape}
conformation is obtained when $\kappa _{s}$ is large.
In this state, the flexible block at the core is surrounded by
the semiflexible block (Fig. \ref{fig:1} (II)).
Intuitively, our core-shell structure is a composite of a globule and toroid. 
This is a typical characteristics of block copolymers: the folded state of this
block copolymer incorporates the characters of both flexible and semiflexible polymers.
This conformation is observed over a wide range of parameters, while at
$\kappa _{s} \gg 0$ and $\epsilon \ll 1$, the semiflexible block is
unable to fold, resulting in a {\it tap-pole} conformation with a long lifetime (Fig. \ref{fig:1} (III)).

The core-shell structure is robust against the change of the length of
blocks.
The flexible block forms spherical shape, and thus, the size $R_f$ is
proportional to
\begin{equation}
 R_f \sim N_f^{1/3}.
\end{equation} 
Since the exponent is much smaller than 1, the size is insensitive to
the length.
Furthermore, the size of the semiflexible block, $R_s$ is
\begin{equation}
 R_s \sim N_s^{1/5},
\end{equation}
which is also insensitive to the length of the block.
In fact, both $N_f=64$ and $N_f=256$ blocks reproduce the core-shell
structure in simulations (data not shown).

The final conformation of the rod-coil copolymer is the uniquely determined core-shell structure,
in which the semiflexible block can only assume a toroidal conformation.
In contrast, a semiflexible homopolymer possesses metastable conformations, such as a cylinder (fig. \ref{fig:1}D).
The toroidal and the cylindrical conformations are both stable, and, over the practical time scale,
the distribution does not change since a transition
between two states does not occur.
We will come back to this point later.

\begin{figure}
\includegraphics{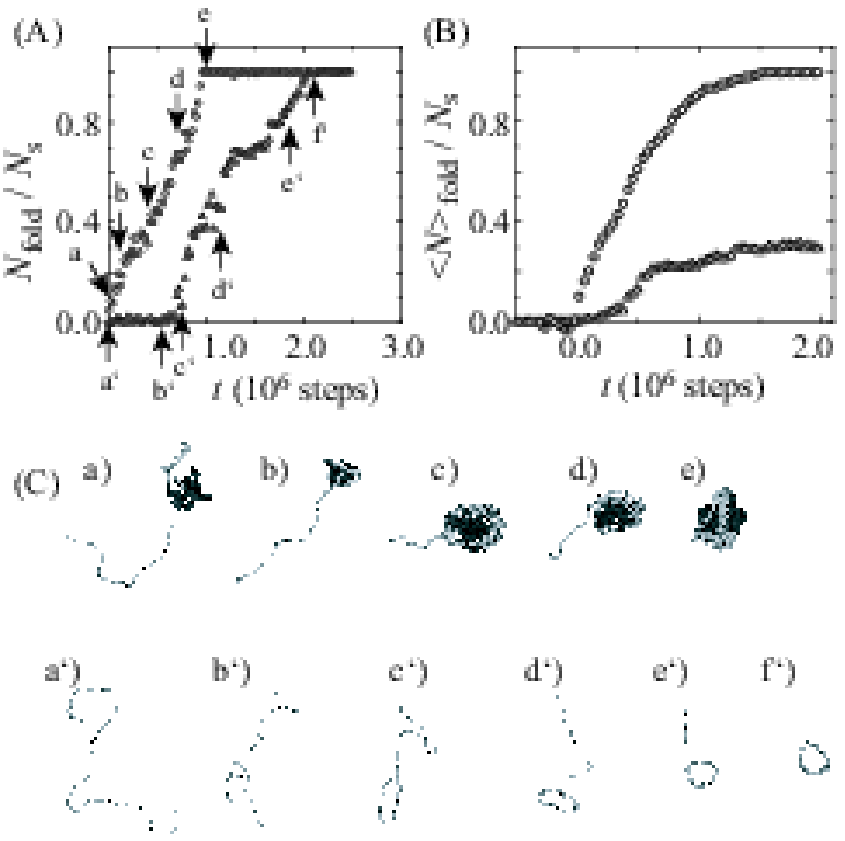}
\caption{Time evolution of the number of collapsed monomers in the semiflexible block
 of a rod-coil copolymer ($\medcirc$) and a semiflexible homopolymer
 ($\bigtriangleup$).
Each polymer has $N=256$ monomers, and a bending elasticity $\kappa _s=15$
 for the semiflexible parts.
We change the attractive interaction at $t=0$ to
 $\epsilon = 0.8$.  
$N_{\mathrm{fold}}$ shows the nuber of monomers in the
 folded state.
The bare data are shown in (A), while the mean evolution over 30 runs is shown in (B).
Snapshots of the polymer conformation during the transition for a
 rod-coil copolymer (a-f) and a semiflexible homopolymer (a'-f') are
 shown in (C).
\label{fig:2}
}
\end{figure}

We show in Fig. 2 the kinetics of the formation for a {\it Saturn-shaped} structure, where
we suddenly increase $\epsilon$ at $t=0$.
The flexible block collapses at an early stage, and the semiflexible
block then gradually wraps around it.
This is contrast to the situation for flexible and semiflexible homopolymers:
flexible polymers exhibit spinodal decomposition 
in the folding process \cite{buguin:1996}, whereas semiflexible polymers
undergo a nucleation and coarsening process \cite{sakaue:2002}.
Due to a stochastic feature of nucleation, there is a long lag time before the folding
transition in a semiflexible polymer takes place.
These observations indicates that a flexible polymer collapses much more
quickly than a semiflexible
polymer.
Figure 2 shows the time evolution in the collapse ratio.
As we can see, rod-coil copolymers undergo folding more quickly than semiflexible homopolymers.
While at $t \sim 350$, a rod-coil copolymer undergoes complete folding on average, only 30 \%
of the monomers in a semiflexible homopolymer exhibit a collapsed states.
The folding of semiflexible homopolymers is subject to a long lag time
originating from the nucleation
process, while rod-coil copolymers start to
fold without such a process due to the existence of the flexible-block core.
As a result, the folding transition proceeds quickly.


These results are confirmed by the consideration of the time scale of
folding in flexible and semiflexible polymers \cite{yoshinaga:2006}.
The initial nucleation time depends exponentially on the persistence
length of the polymer.
This leads to a clear separation of the time scale between the flexible and
semiflexible blocks.
Thus, the folding kinetics of a rod-coil copolymer can be expected to be
a two-step process: folding of the flexible block, and coarsening due to
adsorption of the semiflexible block onto the globule of the flexible block.

\begin{figure}
\includegraphics{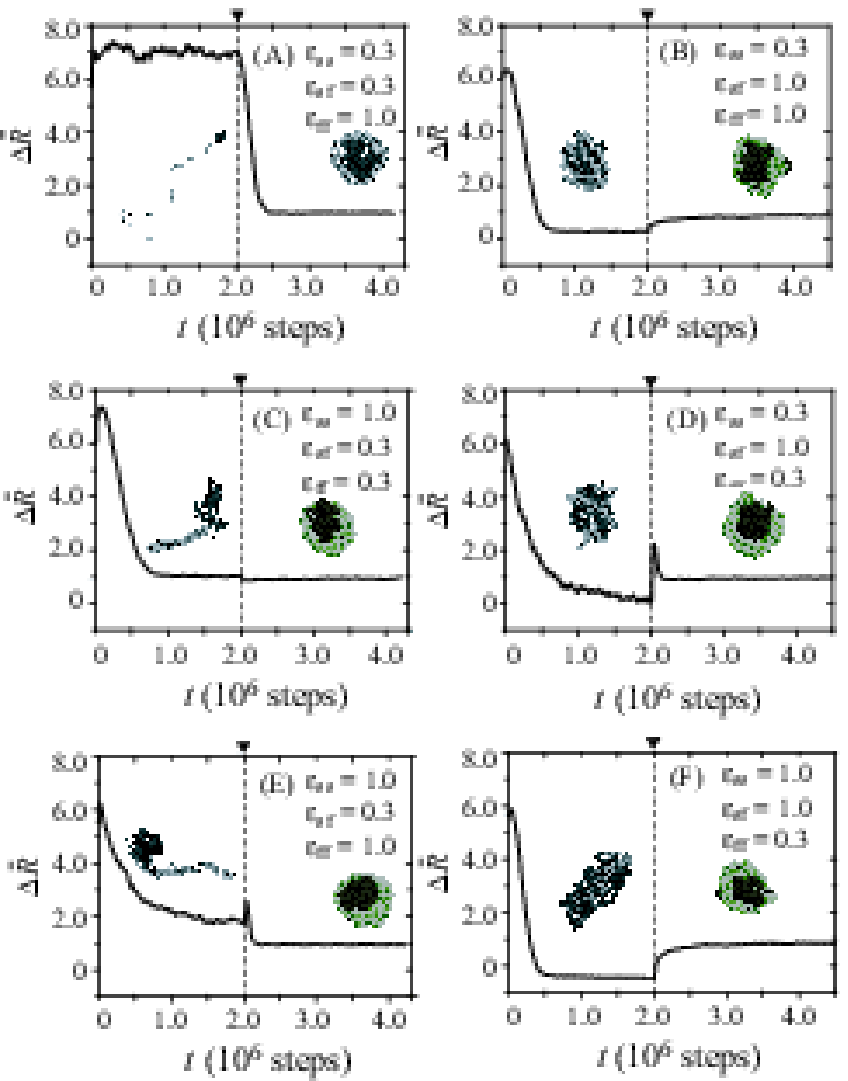}
\caption{Pathways of the folding kinetics of a rod-coil
 copolymer. The effective sizes of the flexible ($\alpha=f$) and a semiflexible
 ($\alpha=s$) blocks,  $\bar{R}_{\alpha}$ are defined as
 $\bar{R}_{\alpha} = \int r \rho_{\alpha} (r) \diff r /
 \int \rho _{\alpha} (r) \diff r$, where $\rho_{\alpha}$ is number density of each block
 in the radial direction. 
Here, $r$ represents the distance from the center of mass. 
The value of $\Delta \bar{R} = \bar{R}_s - \bar{R}_f$, which represents
 the extent to which the flexible block is located outside the
 semiflexible block, is plotted for the various
 intermediate states.
The systems are quenched into
 six intermediate conditions at $t=0$, and then into the final condition
 at $t=2\times 10^{6}$ as indicated by the dashed line. Representations of the initial, intermediate and
 final configurations are shown in the figure. 
The trajectories are averaged over 100 runs.
\label{hetero}
}
\end{figure}

\section{Pathways}
Next, we move our focus to pathways to the folded state.
Although in the above case all the parameters were quenched simultaneously, this is
not always the case.
For this reason, we consider two-step procedures for changing the
parameters, $\{\epsilon
_{ss},\epsilon _{sf},\epsilon _{ff} \}$ from $\{ 0.30, 0.30, 0.30 \}$ to $\{ 1.0,
1.0, 1.0 \}$.
There are six possible intermediate states, as shown in
Fig. \ref{hetero}, which have various intermediate
conformations depending on the
procedures used.
Several of the intermediate states show cylindrical shapes in
their semiflexible parts.
Therefore these structures could be considered to be metastable
states because the core-shell shape mentioned earlier contains a toroidal conformation.   
However, this is not the case.
Interestingly, the final structure does not depend on the procedures used.
This indicates that our core-shell structure is also independent of the
order in which the procedure takes place.
This point is clearly demonstorated in the case of $\{\epsilon
_{ss},\epsilon _{sf},\epsilon _{ff} \} = \{ 1.0, 1.0, 0.30 \}$ (Fig. \ref{hetero}F), where
$\Delta \bar{R} = \bar{R}_s - \bar{R}_f$ changes negative to positive. 
Under these conditions, the intermediate state exhibits a core-shell
structure in which {\it the core is the semiflexible block} and {\it the
shell is the flexible block}. 
As we showed, in the final core-shell structure (Fig. 1(II)) the core is
the flexible block and the shell is semiflexible block; therefore, the
intermediate state shown in Fig. 3(F)
has an {\it inside-out} composition. 
Nevertheless, the core-shell shown in Fig. 1(II) is always obtained as the final structure.

\begin{figure}
\includegraphics{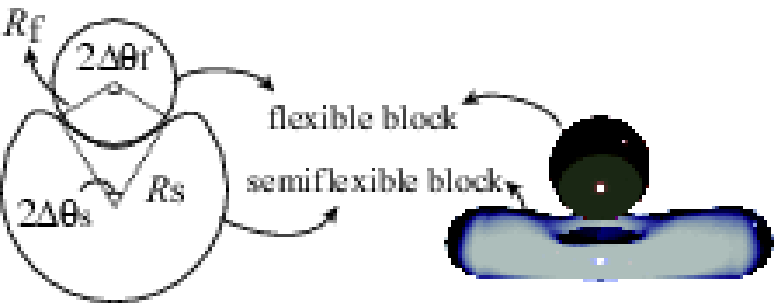}
\caption{ 
Deformation of a cylinder resulting from interaction with a globule
 composed of a flexible block. The left-hand figure shows a
 cross-section of the complex consisting of a spherical globule and a cylinder. 
\label{deformation}
}
\end{figure}

Why is the core-shell conformation independent of the pathway?
First, we should stress that a semiflexible polymer has several
metastable states, including a toroid and a cylinder.
As shown in Fig. \ref{fig:1}D, the conformation of a semiflexible homopolymer
depends on pathway of the transition.
The formation of a cylindrical shape at an early stage of the kinetics leads to a
cylinder at the final state.
The free energy barrier between the states is sufficiently high that the transition
between them does not occur over a practical time scale.
However the addition of a flexible block removes the metastable states.
Hereafter, we discuss the reasons for pathway-independence.
To see the stability of the cylindrical conformation, it is reasonable
to assume the size of the cylinder is of the order of the persistence
length, because cylinders obtained experimentally and in simulations are around this size.
In this case, the conformational fluctuation is too small to undergo a
transition to a toroidal shape when two ends of a cylinder meet.
The only pathway from a cylinder to a toroid is to make a hole at
the center of a cylinder.
We consider a complex of a globule composed of a flexible block and
a cylinder composed a semiflexible block. When the globule is slightly
embedded in the cylinder (Fig. \ref{deformation}), the free energy of
the system due to the deformation,
compared with the reference state, is
\begin{equation}
 \Delta F = F_{\rm int} + F_{\rm sur} + F_{\rm bend}.
\end{equation} 
The reference state is chosen as the conformation in which the globule and the
cylinder are separated, without any interactions.
$F_{\rm int}$, $F_{\rm sur}$ and $F_{\rm bend}$ represent the attractive interaction between
the flexible and semiflexible blocks, and the increase in surface energy and the
bending energy due to deformation.
The energy represented by the first term contributes to deformation,
while those of the second and third terms prevent it.
Since the attractive interaction is short-range, $F_{\rm int}$ is
proportional to the contact area.
This energy per unit area can be represented by $-\epsilon _{sf} /a^2$, and thus, the free
energy is written as
\begin{equation}
 F_{\rm int} \simeq  - \frac{4 \pi \epsilon _{sf} R_f^2}{a^2} (\Delta \theta _f)^2,
\end{equation}
where $\Delta \theta _f$ is the angle characterizing the
deformation (Fig. \ref{deformation}).
The bending free energy at the lowest order in $\Delta \theta _f$ is
proportional to square of the curvature, which is
approximately $\Delta \theta _f/L$, and thus
\begin{equation}
 F_{\rm bend} 
\simeq 
\frac{\kappa _s}{6L} \left( \frac{R_f}{a} \right)^2 ( \Delta \theta _f)^4,
\end{equation}
where $L$ is the length of the longer axis of the cylinder.
With volume conservation
and the condition $R_f \sin \theta _f = R_s \sin \theta _s$, we find
that the
surface free energy is proportional to $(\Delta \theta _f)^4$:
\begin{equation}
 F_{\rm sur} \simeq \frac{\pi}{4} (\Delta \theta _f)^4.
\end{equation}
The fact that $F_{\rm int}$ is dominant for small $\Delta \theta _f$
indicates instability in the cylindrical conformation and an increase in contact area. 
We thus obtain the core-shell conformation as the equilibrium state of
the global
free energy minimum.

\section{Summary and Remarks}
We report novel nanostructures made from single rod-coil block copolymers.
The core-shell structure is found to be
the equilibrium state.
This structure is obtained due to incommensuration of the flexible and
semiflexible blocks:
the segments in semiflexible blocks tend to align in folded states, while
those in flexible blocks tend to be disordered.
This state is the counterpart of the coexistence states found in rod-coil copolymer
blends \cite{hashimoto:1986,liu:1994}. 

The kinetics of the block copolymers are markedly different from those of a semiflexible
homopolymers: in a block copolymer, the flexible blocks tend to collapse
quickly, which means that the semiflexible blocks can collapse without a
nucleation process by making use of the globular collapsed parts of the flexible blocks.
On the other hand, semiflexible homopolymers require a long time to enter the folded state
because the time required for nucleation appears as a lag-time before the collapse.
We also discussed the robustness of the core-shell structure:
formation of the core-shell structure does not depend on intermediate structures.
This was discussed in terms of the instability of cylindrical
shapes.

Although our simple model does not specify real macromolecules to which
it can be applied,
inhomogeneity of elasticity is realized, for example, with
binding ligands \cite{besteman:2007}, and in
mixtures of a helix and a coil \cite{nowak:2004} where the helix is relatively stiff.
In protein folding, it was proposed that secondary
structures are transiently formed at an early stage according to their local preference, independently of native
structures \cite{ptitsyn:1975,qina:2001,myers:2002}.
This leads to inhomogeneity of semiflexibility along a chain.
Thus, we expect that general
features of kinetics at later stage of proteins in this class are to be described with a mesoscopic model
of inhomogeneous bending rigidity.
In fact, our polymer has unique folded structure regardless of difference in pathways.
Mesoscopic models for protein folding are still developing.
It was proposed that 
correlation in sequences of hydrophobicity is relevant to the
core-shell structure and, in fact, they reproduced the structure with
designed sequences  \cite{govorun:2001}.
Our approach focuses instead on semiflexibility along a chain, and we
expect that inhomogeneity in semiflexibility plays a key role for transition kinetics in biomacromolecules.


\begin{acknowledgments}
The authors thank A. Zinchenko and A. Halperin for reading the manuscript.
This work was supported by Japan Society for the Promotion of Science
 (JSPS) under a Grant-in-Aid for Creative Scientific Research (Project No. 18GS0421).
N. Y. would like to acknowledge a fellowship No. 1142 and No. 7662 from the Japan Society for
 the Promotion of Science (JSPS).
Numerical computation in this work was in part carried out at the Yukawa
 Institute Computer Facility.
\end{acknowledgments}

\bibliography{/home/yoshinaga/home/research/yoshinaga,note}

\listoffigures

\end{document}